# Evidence of a bisymmetric spiral in the Milky Way

Charles Francis[1], Erik Anderson[2]
[1]*25 Elphinstone Rd., Hastings, TN34 2EG, UK.*
[2]*360 Iowa Street, Ashland, OR 97520, USA*

*Context*: Because of our viewing point within the Galactic disc, it is extremely difficult to observe the spiral structure of the Milky Way.
*Aims*: To clarify the structure of the Galaxy by re-examination of gas distributions and data from 2MASS; to determine stream memberships among local stars and to show the relationship between streaming motions and spiral structure.
*Methods*: We extend the spiral pattern found from neutral gas towards the Galactic centre using data from 2MASS. We select a population of 23 075 local disc stars for which complete kinematic data is available. We plot eccentricity against the true anomaly for stellar orbits and identify streams as dense regions of the plot. We reconstruct the spiral pattern by replacing each star at a random position of the inward part of its orbit.
*Results*: We find evidence in 2MASS of a bar of length 4.2 ± 0.1 kpc at angle 30 ± 10°. We extend spiral structure by more than a full turn toward the Galactic centre, and confirm that the Milky Way is a two-armed grand-design bisymmetric spiral with pitch angle 5.56 ± 0.06°. Memberships of kinematic groups are assigned to 98% of local disc stars and it is seen that the large majority of local stars have orbits aligned with this spiral structure.

## 1 Introduction

It is straightforward to observe spiral structure in other galaxies but extremely difficult to observe it within the Milky Way because, until the Gaia mission produces results, we will not have accurate distance measurements to individual stars on a Galactic scale, and because evidence from the gas distribution and from star forming regions is unclear. This was recently illustrated by observations from the Spitzer telescope showing that stellar concentrations are not found at the positions where two arms had been thought to be (Benjamin, 2008).

Vallée (1995) collated estimates of the pitch angle from magnetic fields, dust, gas, and stars ranging from 5° to 21°, leading to a corresponding ambiguity in the number of arms. The usual four-armed spiral is derived principally from the distribution of ionized hydrogen (Georgelin & Georgelin, 1976; Russeil, 2003), but in fact the distribution is so sparse and irregular that it is difficult to be certain that anything has really been fitted. Hou, Han and Shi (2009) have collated more recent data to study the distribution of giant molecular clouds (GMCs) and HII regions, and find a number of possible two- and four-armed spirals. However, none of the fits are really convincing, and may be impaired by the difficulty of obtaining accurate distance measurements. In an alternative approach, the neutral hydrogen distribution was famously mapped by Oort, Kerr and Westerhout (1958), and more recently by Levine, Blitz and Heiles (2006). Levine, Blitz & Heiles fit a (somewhat irregular) four-armed spiral, but comment that other fits are possible.

In section 2 we fit a two-armed bisymmetric logarithmic spiral with the hydrogen distribution. We match a two-armed spiral fitted to neutral hydrogen with the distribution of GMCs and HII regions given by Hou, Han & Shi. We apply the method of Benjamin (2008) to data from 2MASS in section 3. 2MASS is a considerably more extensive database than GLIMPSE, used by Benjamin. The 2MASS data confirms a two-armed spiral with a confidence of at least 99.8%, improves the estimate of pitch angle, giving 5.56 ± 0.06°, and shows a bar of length of 4.2 ± 0.1 kpc at an angle of 30 ± 10°, assuming a (scalable) distance to the Galactic centre of 7.4 kpc, consistent with recent determinations (Reid, 1993; Nishiyama et al., 2006; Bica et al., 2006; Eisenhauer et al., 2005; Layden et al., 1996).

The existence of moving groups was first established from astronomical investigations dating as far back as 1869 (Eggen, 1958). Distinct from clusters and associations, stellar streams are all-sky motions. Beginning in 1958, Eggen produced a series of seminal studies of stellar streams. Eggen hypothesized that, as star clusters dissolve during their journeys around the Galaxy, they are stretched into tube-like formations, which were subsequently called superclusters. A wide range of stellar ages was identified within superclusters, challenging Eggen's hypothesis of common origin (e.g., Chereul et al., 1998, 1999). The search for other types of dynamic mechanisms to account for streams has been ongoing. Candidates have included migrations of resonant islands (Sridhar & Touma, 1996; Dehnen, 1998) and transient spiral waves (De Simone et al., 2004; Famaey et al., 2005) in which streams originate from perturbations in the gravitational potential associated with spiral structure.

Section 4 will characterise stellar orbits by eccentricity, $e$, and true anomaly, $\phi$, (the angle between the star and its projected orbital pericentre, subtended at the Galactic centre). Streams are identified by dense regions of the $e$-$\phi$ frequency distribution. Orbits are then extrapolated and found to align with the bisymmetric spiral found from 2MASS and the gas distribution. The majority of disc stars have orbits which are at least loosely aligned with spiral arms, while the tightest alignments are seen for the densest regions of the streams. We summarise our conclusions in section 6.

## 2 Fitting to gas and HII regions

We found a visual fit to the hydrogen maps of Oort et al. and of Levine et al. for bisymmetric spirals with pitch angles in the range

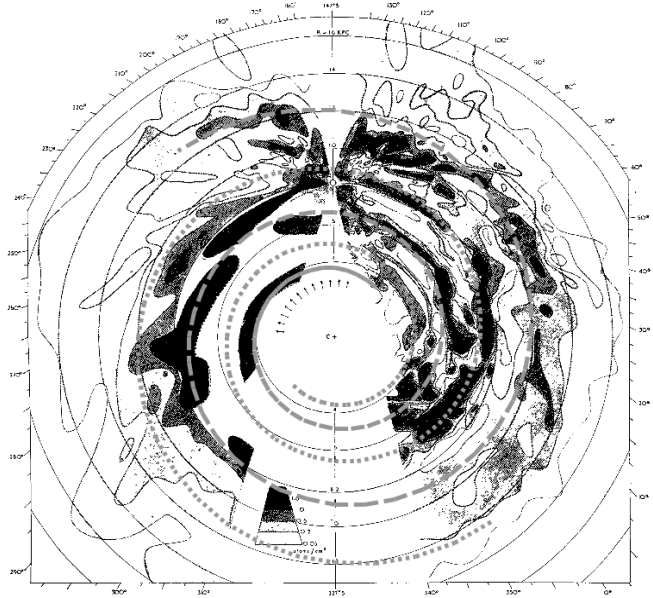

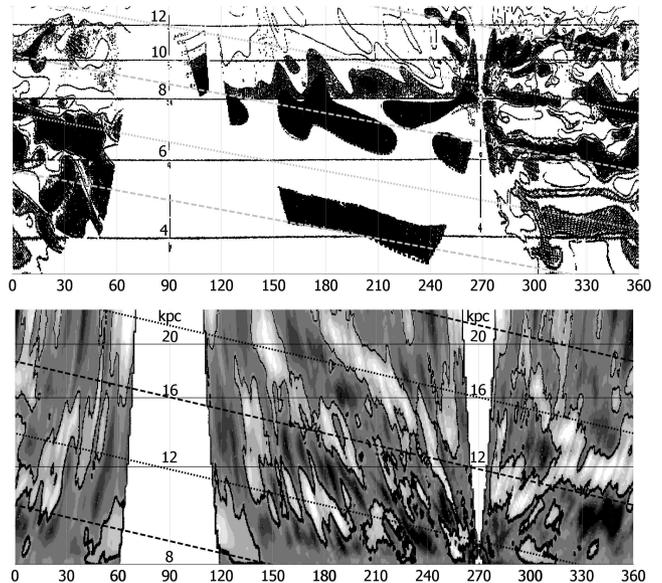

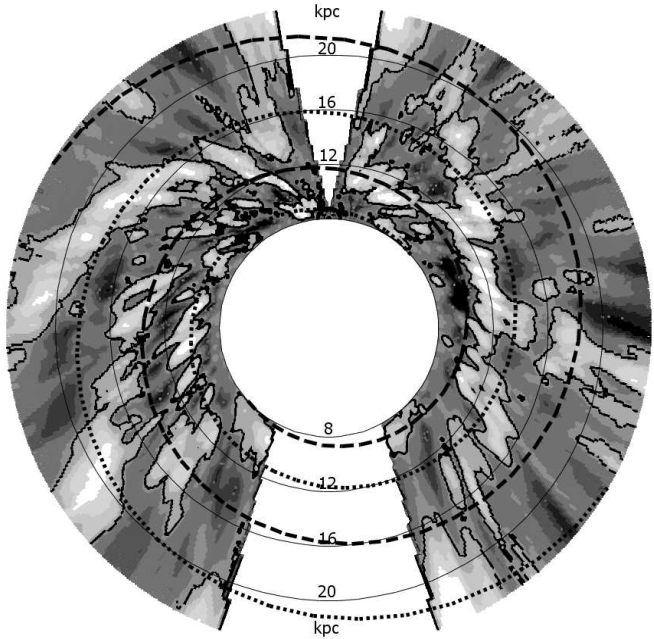

**Figure 1:** Axisymmetric logarithmic double spiral with pitch angle 5.44° fitted to hydrogen maps of Oort et al. and Levine et al. Levine's map has been replotted as a greyscale image, with more dense regions in lighter shades, and shows density relative to a surrounding region, not absolute density. A (scalable) distance to the Galactic centre of 8 kpc has been used for these plots.

5.4 ± 0.5° (figure 1). The range is determined such that the spiral passes through the four dense regions on the right of the Oort map. There is a subjective element in the quality of such a fit, but the two-armed spirals seem to us to better follow the line of the hydrogen clouds, while more open four-armed spirals appear to follow clouds bridging the true line of the arms. The Levine map shows evidence of uncertainties in distance determinations, as dense regions are elongated radially from the Sun. This radial smearing considerably impairs its value as an indicator of spiral structure.

**Figure 2:** The hydrogen maps of Oort et al. and Levine et al. replotted on Cartesian axes, with log($R$) on the vertical axis. The angular scale (horizontal axis) refers to the angle from the Galactic centre, measured clockwise with the Sun at 270°. In these coordinates a logarithmic spiral will appear as equally spaced parallel lines. A bisymmetric logarithmic spiral with pitch angle 5.44° is superimposed, as in figure 1.

By following lines of maximum surface density, Levine et al. found four spirals with pitch angles of 20° - 25°. These are apparent in the region beyond about 12 kpc from the Galactic centre. Two of these are close to symmetrical, while the other two are on the same side of the galaxy. It does not appear to us that high pitch angles continue inward and this may be an indication that spiral structure is looser in the outer regions of the Milky Way, as is not unusual among other spiral galaxies.

Because of the ragged nature of the gas distribution, numerical fitting methods are problematic, and do not appear to us to be better than visual fitting; small deviations from symmetry between the arms and variations in pitch angle may cause an exact logarithmic spiral to fall into a trough between the arms, leading to false results, and because density is not independent for points sampled along a particular path it is not possible to quantify a goodness of fit. The method followed by Levine also suffers because the shorter absolute length of arms with greater pitch angle, together with radial smearing of gas clouds, increases the likelihood of a false fit.

A logarithmic spiral appears as equidistant parallel straight lines when log($R$) is plotted against the angle subtended at the Galactic centre (figure 2). The replotted Oort map (top) shows strong indications of a bisymmetric spiral, and an absence of other reasonable candidates. The replotted Levine map is less clear, but dense regions appear to lie on a bisymmetric spiral, with an increasing pitch angle at greater radii. We observe no sign of another regular pattern. The candidates for arms identified by Levine have an irregular structure which does not continue toward the inner part of the Galaxy.

Pitch angles in the range 5.4 ± 0.5° give good agreement with 5.1° and 5.3° found from HII regions and giant molecular clouds for the two-armed logarithmic model by Hou, Han and Shi (2009). It is observed that, while parts of Hou et al.'s map show a very good fit with the bisymmetric spiral, GMCs and HII regions give a less good fit to any spiral pattern than the HI distribution (figure 3). This may be explained because Baba et. al. (2009, and references cited therein) have shown from a range of VLSI observations that star forming regions and young stars often have large peculiar motions. As a result, distance estimates based on the rotation curve are inaccurate. Moreover, because of their motion with respect to the spiral, GMCs and HII regions may be less good tracers of spiral structure than clouds of neutral gas. Nonetheless, in the plot of log($R$) against angle subtended at the Galactic centre, there is evidence of tracers for spiral arms with pitch angle ~5.5°, and tracers at other angles are not seen.

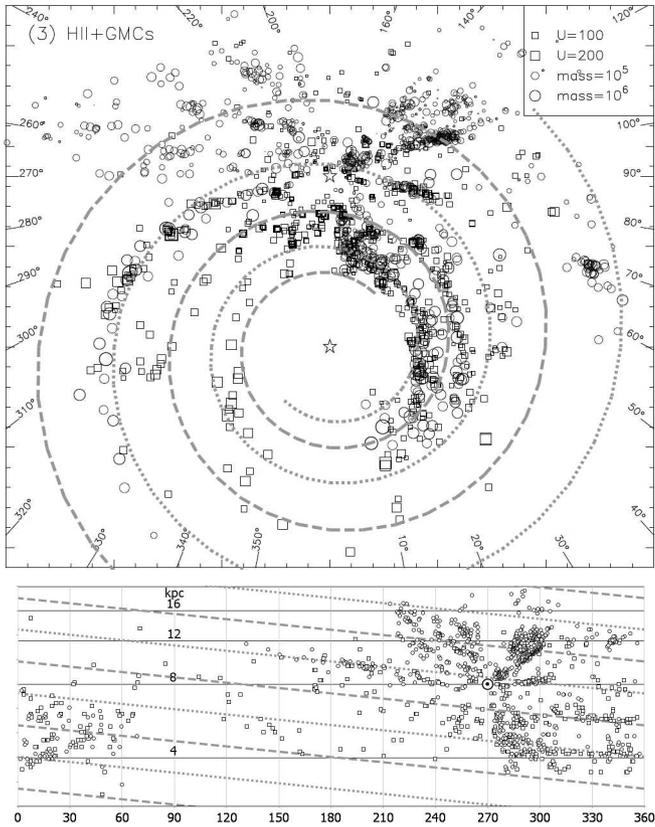

**Figure 3:** The distribution of GMCs and HII regions from Hou, Han and Shi (2009), with the spiral structure found from neutral gas superimposed. The lower plot shows log($R$) against angle subtended at the Galactic centre, measured clockwise with the Sun at 270°.

## 3  Fitting to 2MASS

Benjamin (2008) counted the density of sources from the 6th to 12th magnitude using the GLIMPSE database (a compilation of Spitzer and 2MASS data). It is to be expected that higher counts will be found in directions tangential to the arms. Benjamin established that major spiral arms are not in the positions predicted by four-arm models. Benjamin's clearest results came from counts of the 2MASS J, H, and K bands. They were restricted to Galactic longitudes |$l$| < 70° and to magnitudes brighter than 12 because he used only sources also detected by Spitzer. Since 2MASS is an all-sky survey and contains stars to around the 20th magnitude, we removed these restrictions by working directly from the 2MASS data (Cutri et al., 2003).

We restricted 2MASS to stars with *use* = 1, (use source flag) *Cflg* = 0 (contamination and confusion flag) and with Galactic latitude |$b$| < 1°. Using 1° bins in Galactic longitude we made counts of all sources, and counts of sources from the 6th to 15th magnitude in the mean of the J, H and K bands (for which signal to noise ratios are generally in the order of at least 10). We counted valid measurements in each band (as determined from $Qflag$ = "A", "B","C","D"), but did not make restrictions on signal to noise ratio as this might introduce a selection effect reducing counts in dense regions.

We plotted the frequency distribution (figure 4, upper plot). For the number of stars in the population, random variations in the number of stars in each bin will be below 400 – a couple of orders of magnitude below a visible change in the vertical axis. Thus every visible peak and trough in the frequency distribution is evidence of structure (including spurs and clouds, as well as spiral arms and the bar). We matched peaks in the observed frequency distribution to putative positions of tangencies to the spiral arms and the bar (lower plot), finding a remarkably good match with the spiral found from the distribution of neutral gas going down to a much lower than expected Galactic radius, and a short bar.

The correspondence between peaks in the frequency distribution and tangencies to a bisymmetric logarithmic spiral is much better than one would expect given the variety and imperfections of other spiral galaxies. Each of nine tangencies to the arms from the Solar position lies precisely on a peak (there are some peaks which are not well represented in the plot of all sources, but these are seen clearly in the plot of 6-15 mag sources). It is highly improbable that this degree of correspondence could have come about by chance. For a random distribution, an order of magnitude estimate of the probability for nine tangencies to lie on peaks is less than 0.2%, assuming a 50% probability that each one lies on a peak. Since this substantially overestimates the true probability for a tangency to lie on a peak of the distribution, one can reasonably estimate a probability at least two orders of magnitude smaller.

The tangencies account for most of the main features of the frequency distribution, but there are also notable spurs at $l$ = 88° and $l$ = 48° and lesser spurs at $l$ = 36° and $l$ = –32°. Data counts for all sources also increase at $l$ = –48°, but this is barely visible in the 6th-15th magnitude count, suggesting that there may be a spur outside the Scutum Crux arm. A region of extinction splits the tangency of the Scutum arm in the 6th-15th magnitude count, but has minimal impact on the count of all sources.

The bar can be expected to consist of stars in highly elongated orbits aligned on an axis. We therefore expect a peak in the distribution near the Galactic centre, where stars in both arms of the bar are close to pericentre, and the population overlaps with core stars with low orbital radii. Moving out from the centre, the first expected peaks in the velocity distribution should be the ends of the bar, where the stellar density is greater because stars are moving relatively slowly close to apocentre, and observed density is also increased because of the tangencies of orbits near the end of the bar. The correct identification of the bar is confirmed because of the asymmetry between the first peaks to either side of the central position; source counts in the further end of the bar are markedly lower than those in the near end, whereas the rest of the distribution is more symmetrical.

Figure 5 shows the positions of the ends of the bar, as seen in the 2MASS data, on the central regions of images from COBE (Boggess et al., 1992) and WISE (Wright et al., 2010). This corresponds closely to the asymmetry seen particularly in the COBE image, such that the visual width of the Galaxy is greater at the near end of the bar, and confirms the boxy structure, or bulge, described by e.g. Dwek et al. (1995). The size of the first peaks to either side of the core indicate substantial tangencies, suggesting that the bulge has an approximately elliptical cross section, as seen in numerous face-on galaxies. The degree of assymetry between the first peaks is consistent with a bar length of $4.2 \pm 0.1$ kpc at an angle of $30 \pm 10°$, and with axis ratio $1:0.4 \pm 0.05$.

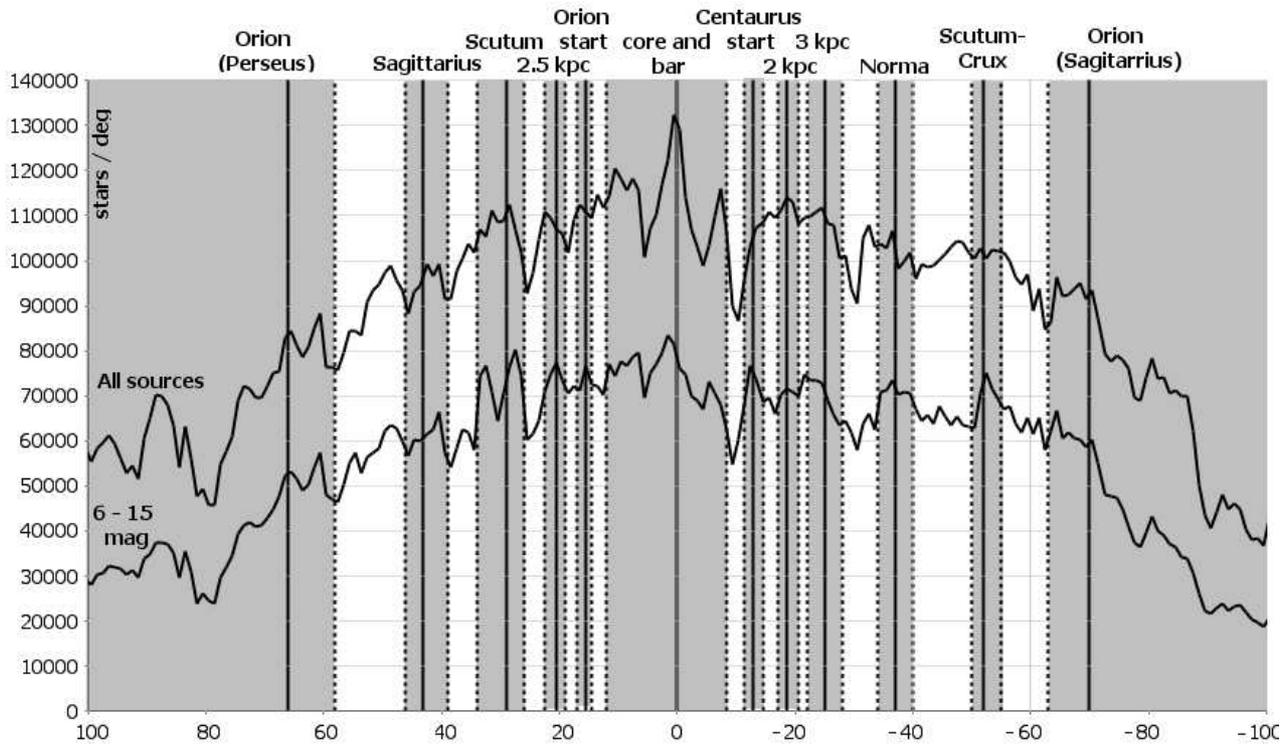

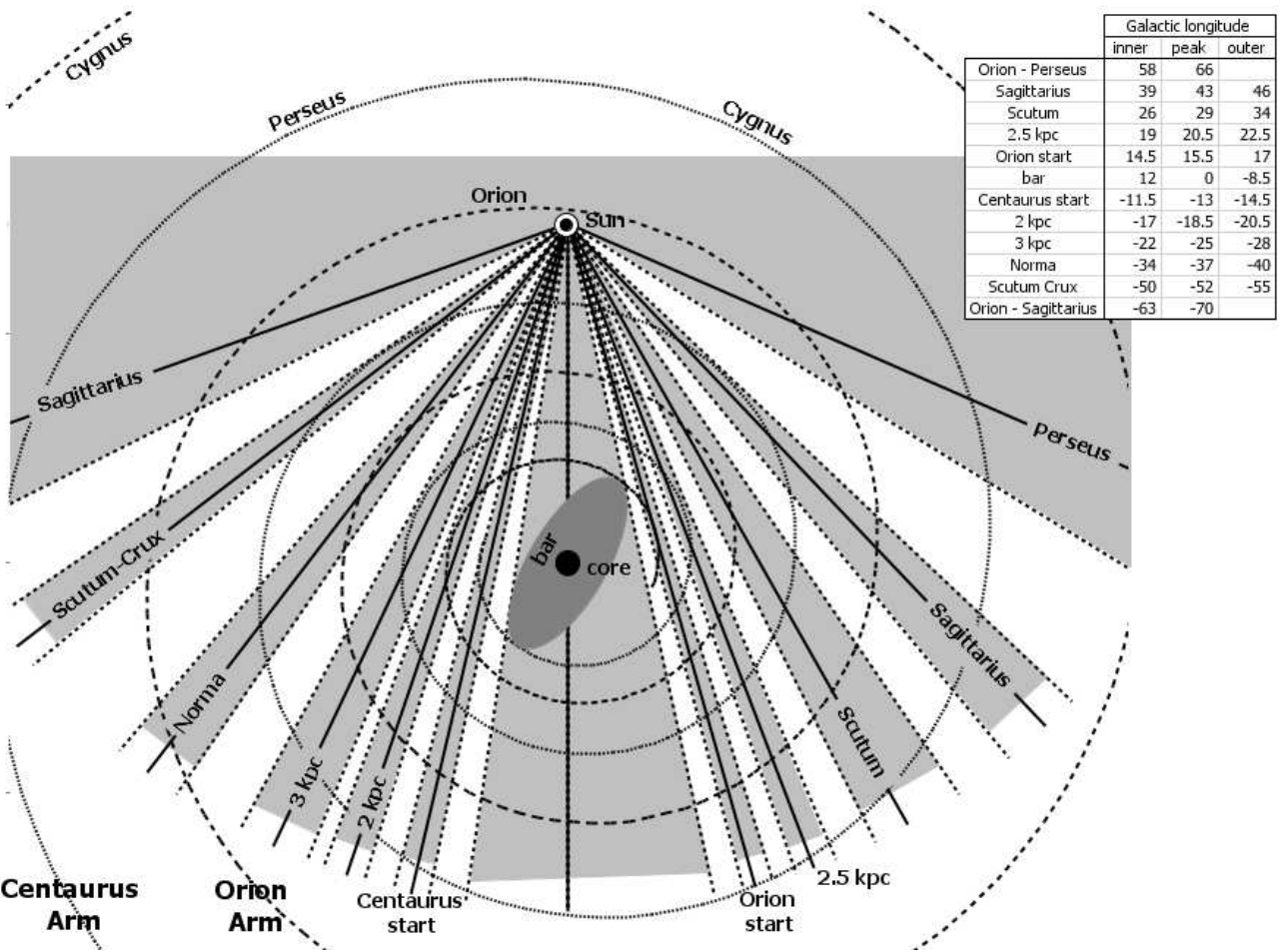

**Figure 4:** The correspondence between the frequency distributions for all sources in 2MASS with galactic latitude |*b*| < 1°, and sources from the 6th to 15th magnitude (upper plot) with tangencies to spiral structure and the bar (lower plot). The mean magnitude of the J, K H bands was used. The bar is seen in the asymetric peaks to either side of the core. The table shows plotted values of Galactic longitude.

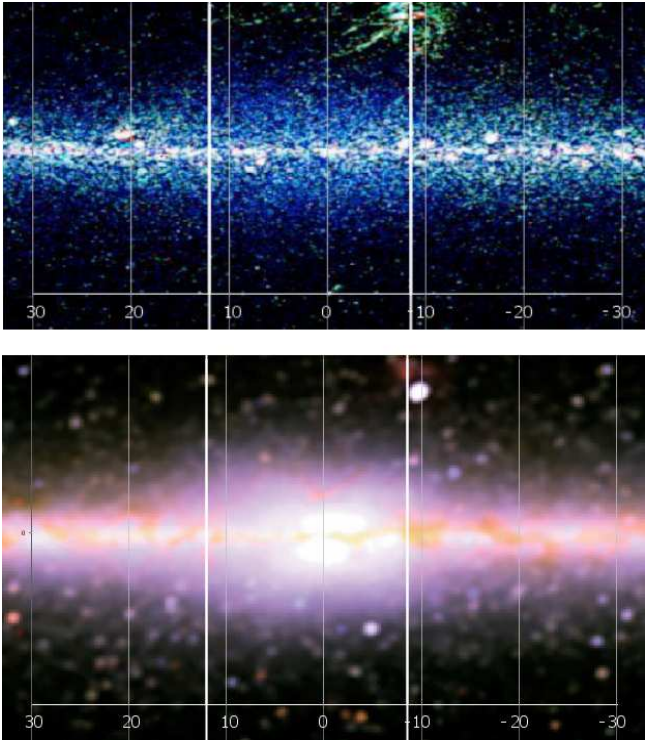

**Figure 5:** Central regions of DIRBE 1.25, 2.2, 3.5µm composite image from COBE (top) and WISE image (bottom) with the end positions of the bar, as seen in the 2MASS data, superimposed as vertical white lines. The horizontal axis shows galactic longitude in the plane of the disc.

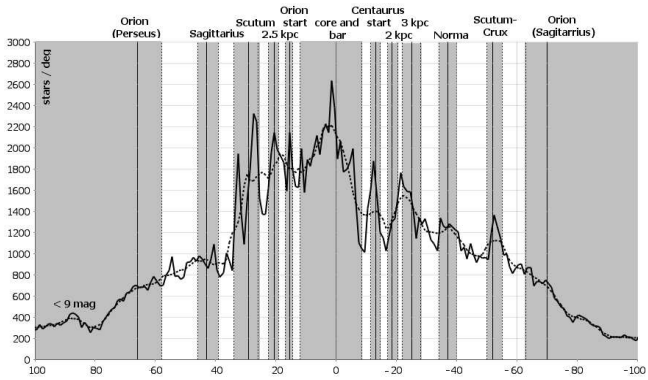

**Figure 6:** The distribution of bright stars, less than 9 mag. The dotted line shows the seven point moving average. Shading denotes the positions of the arms suggested in figure 4.

Restricting the counts to stars with magnitudes less than 9 shows the distribution of recent star formation in the arms (figure 6). There is very little activity in the Sagittarius sector, and a misalignment in the 2 kpc and 3 kpc sectors, but evidence of substantial star formation in Scutum, 2.5 kpc, Orion start, Centaurus start, Norma and Scutum sectors. There is an underlying asymmetry in the distribution of bright stars, seen in the seven point moving average (dotted line). This may be partly explained because tangencies to the arms are closer for positive Galactic latitude, but most likely reflects randomness in the distribution of star forming regions.

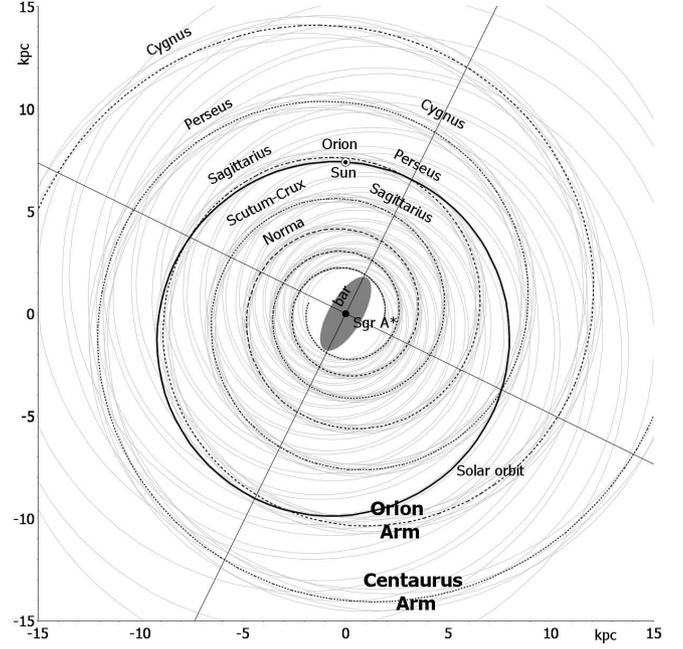

**Figure 7:** Axisymmetric logarithmic double spiral with pitch angle 5.56° constructed from ellipses of eccentricity 1.6. The plot uses 7.4 kpc as the distance to the Galactic centre, but figures are scalable. The solar orbit is shown in approximation (neglecting precession), together with its major axis and latus rectum, as calculated in XHIP.

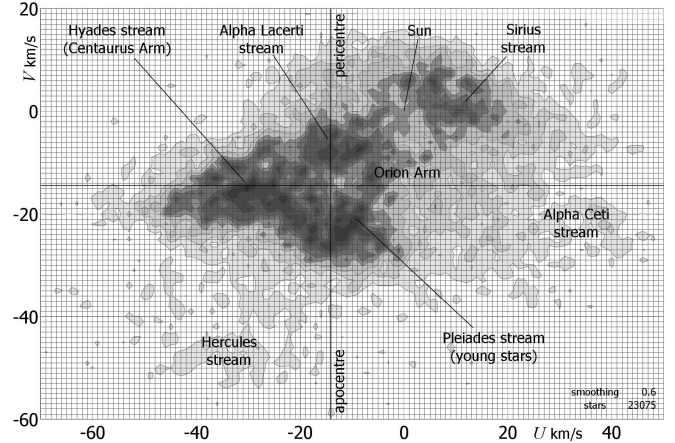

**Figure 8:** The distribution of $U$- and $V$-velocities using Gaussian smoothing with a standard deviation of $0.6 \, \text{km s}^{-1}$, showing the Hyades, Pleiades, Sirius, Hercules, Alpha Lacertae and Alpha Ceti streams. Axes are shown with an origin at the LSR, $(U_0, V_0) = (-14.2, -14.5)$ km s$^{-1}$, as calculated in XHIP.

As is common in spiral galaxies, the start of the spiral is not aligned with the bar. From the position of the inner most peak of the distribution on the Orion arm, and an estimate that the Sun is $200 \pm 100$ pc inside the centre of the arm, we calculate a mean pitch angle $5.56 \pm 0.06°$, in excellent agreement with the pitch angle found from the gas distribution.

The data for $l < -100°$ and $l > 100°$ (not plotted) shows a number of peaks in the count for all sources, but far less structure in the 6th-15th magnitude count. This reflects the notion that star formation (and hence the distribution of bright stars) is more sporadic in the outer regions of the galaxy. We found little or no correspondence between peaks in the distribution and features seen on the maps of either Levine et al. or Hou et al.

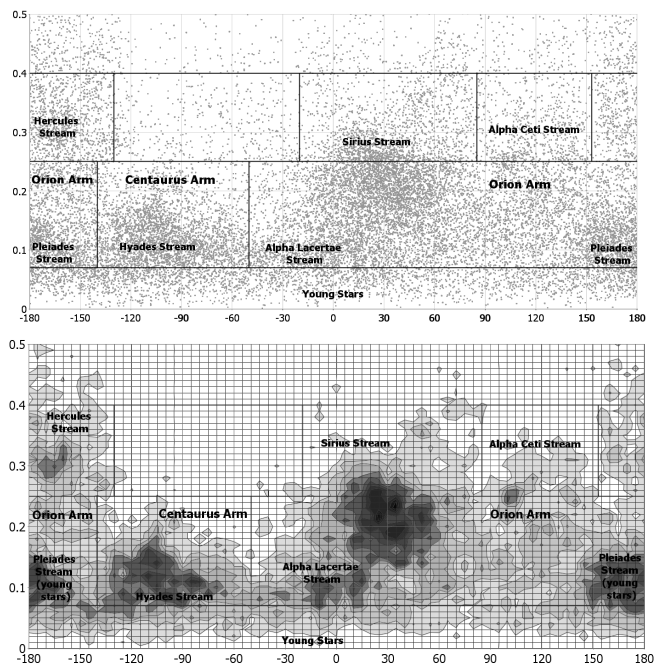

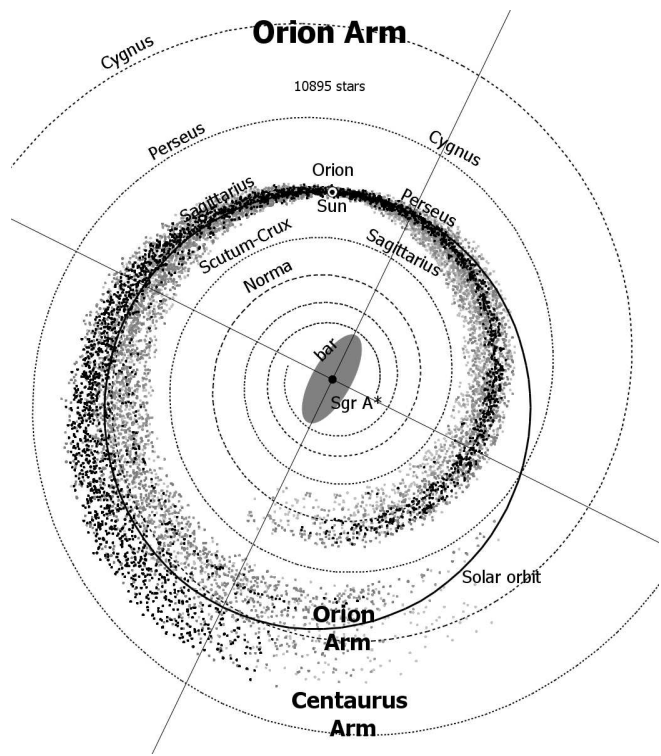

**Figure 9:** a) The distribution of eccentricities, *e,* against true anomaly, φ. Stellar streams are more clearly differentiated in this plot. b) The distribution of eccentricities, *e,* against true anomaly, φ using Gaussian smoothing.

**Figure 10:** Two-armed spiral with a pitch angle of 5.56°, showing the solar orbit (eccentricity 0.159). Orion stream stars are shown at a random position on the inward part of the orbit. Stars in with quality index 1 for stream membership are shown in black, stars with quality indices 2 - 4 are have paler grey for a higher index.

## 4  Fitting to the local velocity distribution

A two-armed spiral necessitates a little care to avoid confusion in naming the arms, because traditionally named sectors with the same name lie on different arms (figure 7). The Sun lies in the Orion sector. This is not a separate spur, but is a part of a major arm connecting Perseus in the direction of rotation to Sagittarius in the direction of anti-rotation. We have called this major spiral arm the Orion Arm. The Orion arm contains Norma, Perseus, Orion, Sagittarius, and Cygnus sectors. The Centaurus arm contains Sagittarius, Scutum-Crux, Cygnus, and Perseus sectors.

The Extended Hipparcos Compilation ("XHIP", Anderson & Francis, 2012) gives radial velocities for 46 392 Hipparcos stars, together with multiplicity information from the *Catalog of Components of Double & Multiple Stars* (Dommanget & Nys, 2002) and *The Washington Visual Double Star Catalog*, version 2010-11-21 (Mason et al., 2001-2010). We limited the database to 23 075 local disc stars, by removing stars outside 300 pc, stars with parallax errors greater than 20%, stars with quoted radial velocity errors greater than 5 km s$^{-1}$, stars with quality index "D" (which includes unsolved binaries), secondary stars in multiple systems, stars in moving groups, and stars with velocities perpendicular to the Galactic plane greater than 24 km s$^{-1}$.

We plotted the velocity distribution for the population (figure 8), using Gaussian smoothing (see e.g. Francis and Anderson 2009) with standard deviation 0.6 km s$^{-1}$. The larger population, and removal of moving groups (especially the Hyades cluster), gives a clearer image of the features of the distribution than has previously been available. In particular there is a clearly defined, and unique, central well at $(U, V) = (-12.5, -14)$ km s$^{-1}$, close to the LSR (local standard of rest, or idealised velocity of circular motion) found in XHIP, $(U_0, V_0, W_0) = (-14.2 \pm 1.0, -14.5 \pm 1.0, -7.1 \pm 0.1)$ km s$^{-1}$. The Hyades, Pleiades, Sirius, Hercules, Alpha Lacertae and Alpha Ceti streams are clearly visible, although there is some overlap between streams.

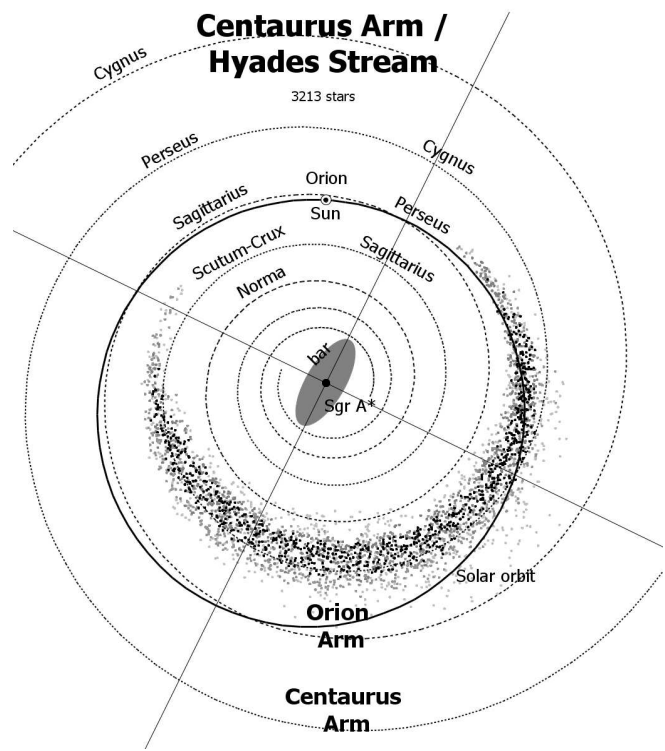

**Figure 10:** As figure 10, but showing stars in the Centaurus arm / Hyades stream.

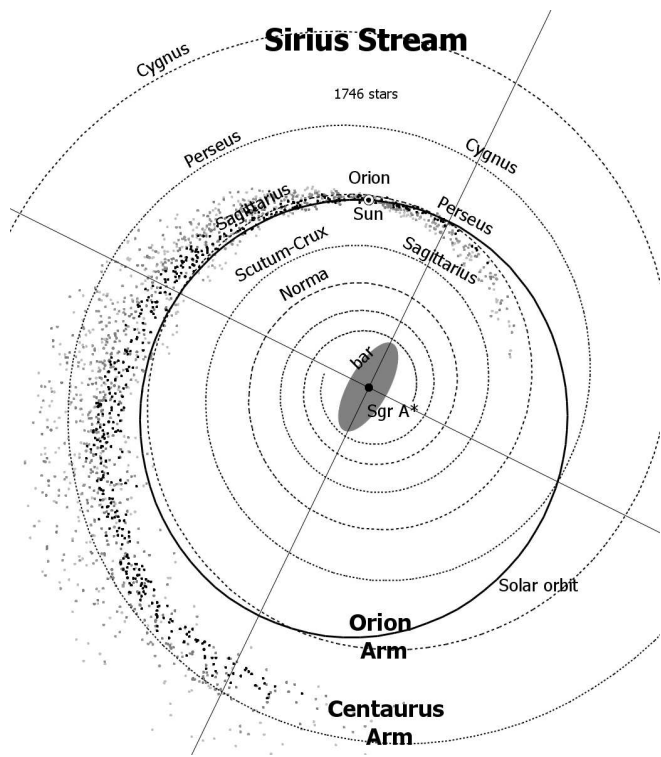

**Figure 11:** As figure 10, but showing stars in the Sirius stream.

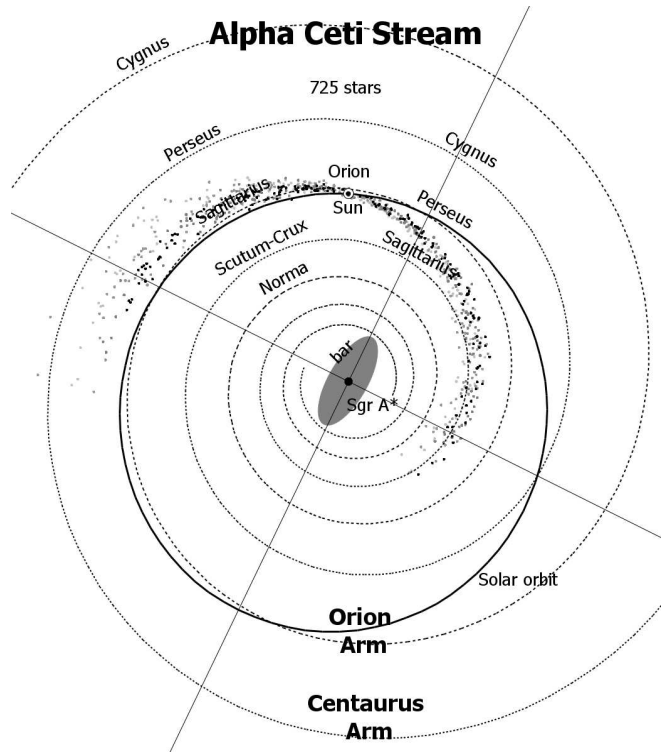

**Figure 14:** As figure 10, but showing stars in the Alpha Ceti stream.

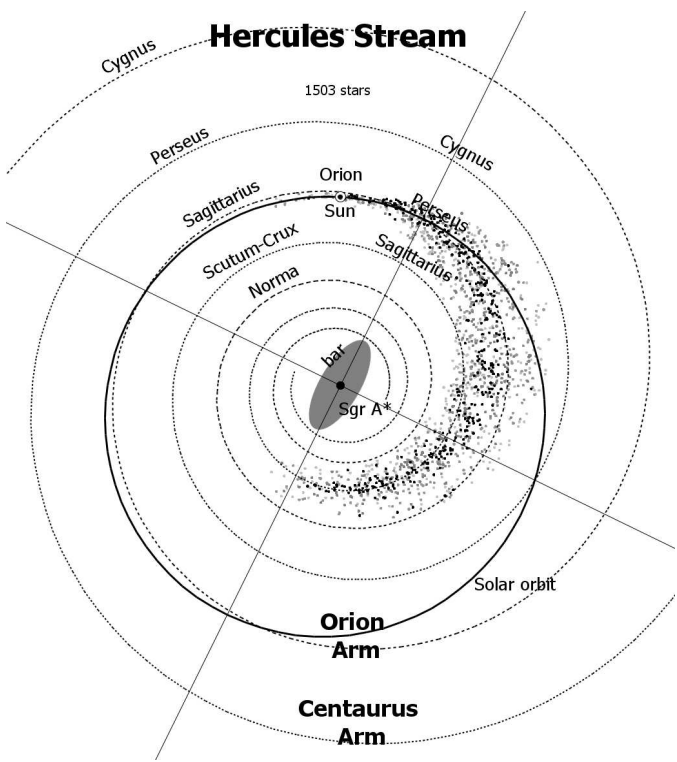

**Figure 13:** As figure 10, but showing stars in the Hercules stream.

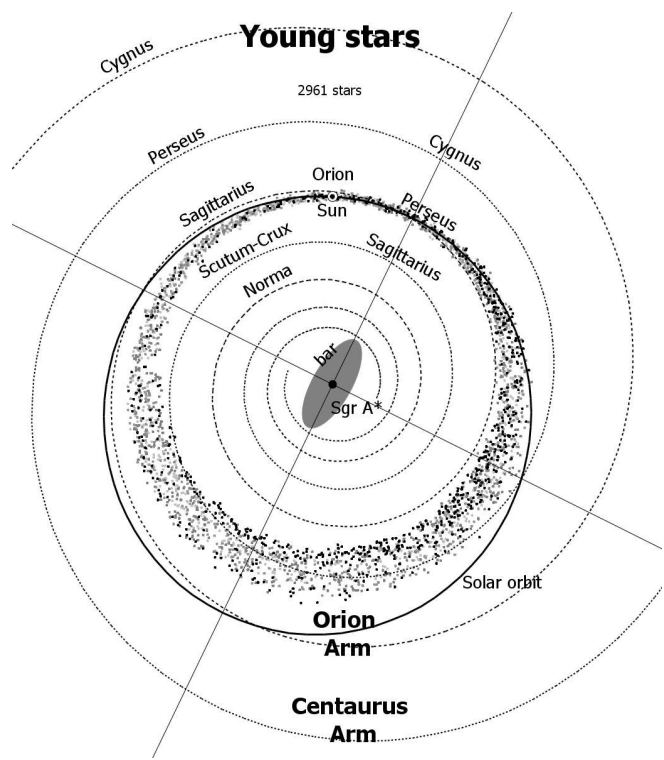

**Figure 15:** As figure 10, but showing young stars.

For an elliptical orbit the eccentricity vector is defined as the vector pointing toward pericentre and with magnitude equal to the orbit's scalar eccentricity. It is given by

$$e = \frac{|v|^2 r}{\mu} - \frac{(r \cdot v)v}{\mu} - \frac{r}{|r|}, \qquad (4.1)$$

where *v* is the velocity vector, *r* is the radial vector, and $\mu = GM$ is the standard gravitational parameter for an orbit about a mass *M* (e.g. Arnold, 1989; Goldstein, 1980). For a Keplerian orbit the eccentricity vector is a constant of the motion. Stellar orbits are not strictly elliptical, but rosette orbits can usefully be regarded as precessing ellipses and the eccentricity vector remains a useful measure (the Laplace-Runge-Lenz vector, which is the same up to a multiplicative factor, is also used to describe perturbations to elliptical orbits).

For local disc stars, over some time span, the orbit can be approximated by an ellipse, characterised by current distance, *R*, to the Galactic centre, eccentricity, $e = |e|$, and the true anomaly, $\phi$, or angle between the eccentricity vector and the star, subtended at the Galactic centre. We plotted the distribution of eccentricities against true anomalies ($e$-$\phi$) for the population (figure 9a), based on an adopted Solar orbital velocity of 225 kms$^{-1}$. We found a much improved differentiation between streams on this plot. We segmented the diagram into subpopulations:

| eccentricity | true anomaly | group |
|---|---|---|
| $e \le 0.07$ | | young stars |
| $0.07 < e \le 0.25$ | $\phi \le -140$ or $-55 < \phi$ | Orion arm |
| $0.07 < e \le 0.25$ | $-140 < \phi \le -55$ | Centaurus/Hyades |
| $0.25 < e \le 0.4$ | $\phi \le -130$ or $153 < \phi$ | Hercules stream |
| $0.25 < e \le 0.4$ | $-20 < \phi \le 85$ | Sirius stream |
| $0.25 < e \le 0.4$ | $85 < \phi \le 153$, | Alpha Ceti stream |
| $0.4 < e$ | | high eccentricity |

These designations assign kinematic group memberships to 98% of local disc stars. They are justified by the alignments seen in figure 10 to figure 14. Nearly half (47.2%) of the local population, including the Sun, have orbits broadly aligned with the Orion arm. 13.9% belong to the Hyades stream and have orbits aligned with the Centaurus arm (figure 10). The Sirius, Hercules and Alpha Ceti streams respectively contain 7.6%, 6.5% and 3.1% of the population, and also consist of stars with orbits aligned with spiral structure (figure 11, figure 13, figure 14). The high eccentricity group contains 6.8% of the population, and young stars with low eccentricity orbits account for 12.8%.

We found the density of the distribution using Gaussian smoothing (figure 9b). We divided the stars in each segment by quartiles of the density at the position of each star on the $e$-$\phi$ diagram, and assigned a quality index, 1 - 4, to each star. Thus, stars in a region with density greater than the upper quartile show the tightest adherence to stream motions, and are given quality index 1, while those in a region less dense than the lower quartile are least matched to stream motions and are given quality index 4.

For each star in each subpopulation we overplotted a random position on the inward part of the orbit, approximated by an ellipse. Stars in with quality index 1 for stream membership are shown in black, stars with quality indices 2 - 4 have paler grey for a higher index. Thus, in figure 10 to figure 15 stars are displaced from their true position by less than half an orbit. For typical orbits this less than about 150 Myrs, much less than stellar ages except in the case of young stars (figure 15) whose orbits do not align with spiral structure. The method neglects orbital precession. If orbital preces-sion were large over half an orbit, then there would be poor alignment with the spiral pattern, but in fact the alignment is good and we can conclude that orbital precession (and spiral pattern speed) is slow.

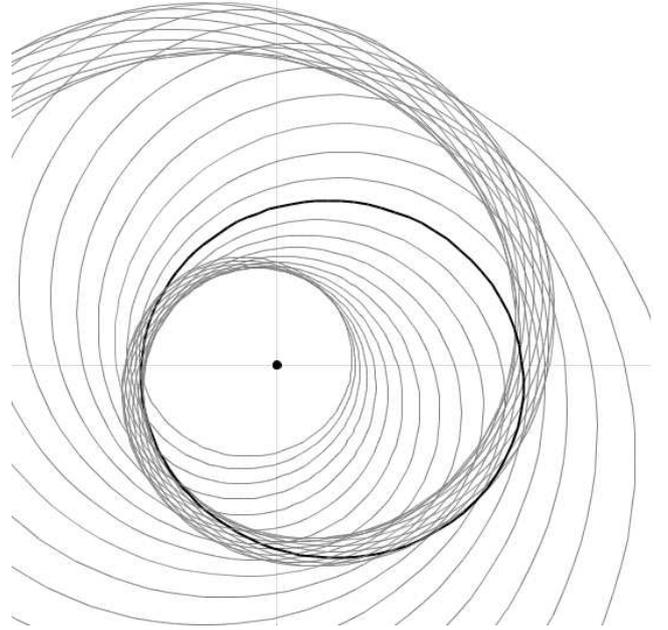

**Figure 16:** An equiangular spiral with a pitch angle of 11°, constructed by repeatedly enlarging an ellipse with eccentricity 0.3 by a factor 1.05 and rotating it through 15° with each enlargement. Ellipses with eccentricity greater than about 0.25 have more than half their circumference within the spiral region. Ellipses with eccentricity greater than about 0.35 produce probably too broad a structure to model a spiral arm with this pitch angle, but give a good fit for spirals with greater pitch angle. Lower eccentricity ellipses produce a narrower spiral structure and/or a fit with spirals of lower pitch angle.

It is seen that the motions of stars in the dense regions of the streams, shown in black, conform most closely to spiral structure, and that the large majority of stars in the solar neighbourhood show some adherence to the arms. We might have hoped to confirm the position of the LSR from the quality of the fit between streams and spiral structure, but in practice the fit is remarkably insensitive to varying the LSR.

The alignments seen in figure 10 and figure 10 can be understood in terms of an orbital model constructed by repeatedly enlarging an ellipse by a constant factor, *k*, centred at the focus and rotating it by a constant angle, $\tau$, with each enlargement (figure 16). The pitch angle of the spiral depends on *k* and $\tau$, not on the eccentricity of the ellipse, but, for a given pitch angle, ellipses with a range of eccentricities can be fitted to the spiral, depending on how narrow one wants to make the spiral and what proportion of the circumference of each ellipse one wants to lie within it. Higher eccentricity orbits fit spirals with higher pitch angles. Thus stars move along an arm on the inward part of their orbits, leaving the arm soon after pericentre, crossing the other arm on the outward part and rejoining the original arm before apocentre. In this description of material arms there is no winding problem because the spiral depends on the *paths* of stellar orbits, not on orbital velocity.

Thus stellar streams in the solar neighbourhood, can be understood as facets of spiral structure. The Orion arm (figure 10) contains stars in the Pleiades, Alpha Lacertae, Alpha Ceti and Sirius streams. The low density in the range ~60 < φ < ~150 (figure 9) results because in this part of the orbit stars tend to the outside of the spiral arm, whereas the Sun is approaching pericentre, and is nearer the inside of the arm. The Alpha Lacertae stream and part of the Sirius stream consist of stars similarly close to pericentre and also occupying the inner part of the spiral arm, while the Pleiades stream consists in part of young stars whose orbits have not yet settled, and in part of stars crossing the inner part of the arm as they approach apocentre.

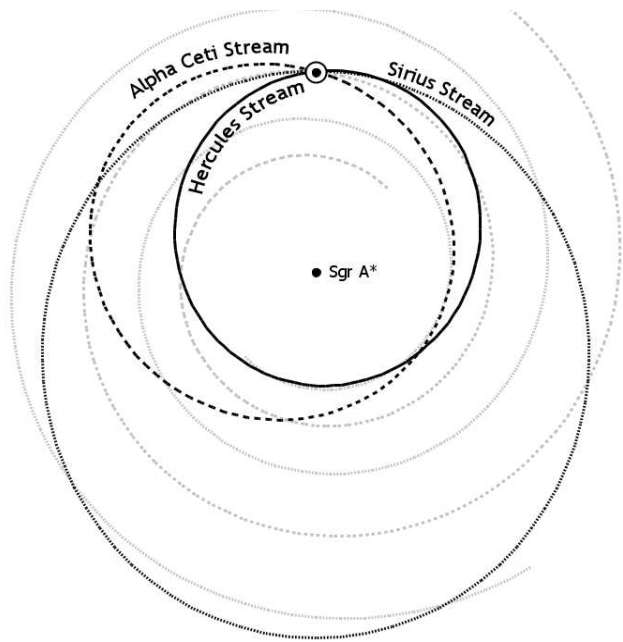

**Figure 17:** Three alignments with spiral structure for stars passing through the locality of the Sun in orbits with eccentricity 0.29. The continuous line shows orbits in the Hercules stream. The dashed line show orbits in the Alpha Ceti stream, and the dotted line represents orbits in the Sirius Stream.

The Hyades stream (figure 10) consists of stars in orbits aligned with the Centaurus arm, as they cross the Orion arm on the outward part of their orbits. Figure 11, figure 13, and figure 14 show stars in Sirius, Hercules and Alpha Ceti streams, corresponding to the alignments with spiral structure shown in figure 17. The presence of the Sirius stream in the local velocity distribution is indicative that the Milky way spiral continues to at least about 15 kpc from the Galactic centre, while the Hercules stream has a radius ~4 kpc at pericentre, and indicates spiral structure continues inwards a least to this radius (the 2MASS data shows it continues inwards for another full turn).

Famaey et al. (2005) identified a kinematic group of young giants with velocities close to that of the LSR. The bulk of stars are created either in low eccentricity orbits (figure 15) or in the Pleiades stream, but there is also some overlap for young stars with the Hyades stream. This can be seen in at the ends of the arms in figure 10, where stars in low eccentricity orbits lie inside the arm at the outer end of the plotted region, and outside the arm and the inner end.

## 5 Discussion

According to the stellar migration hypothesis (Lépine et al. 2003, Roškar et al., 2008) stars do not remain on circular orbits, but are perturbed by spiral structure such that orbital radius varies by about 2-3 kpc over periods in the order of 1 billion years. This prediction has been confirmed by the deep well seen in the velocity distribution at the position of circular motion (figure 8), and by the eccentricities seen in the local population (figure 9).

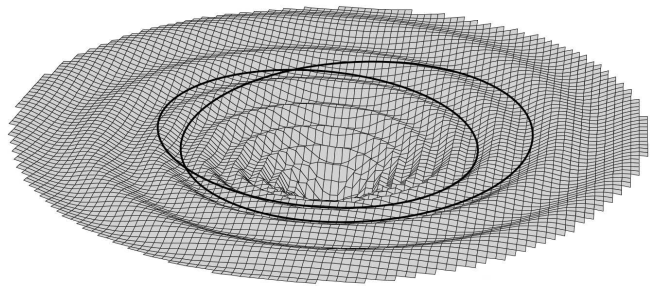

**Figure 18:** The gravitational potential of a bisymmetric spiral galaxy plotted on a vertical axis against the galactic disc on a horizontal plane. The alignment of (idealised) elliptical orbits with troughs in the potential is shown.

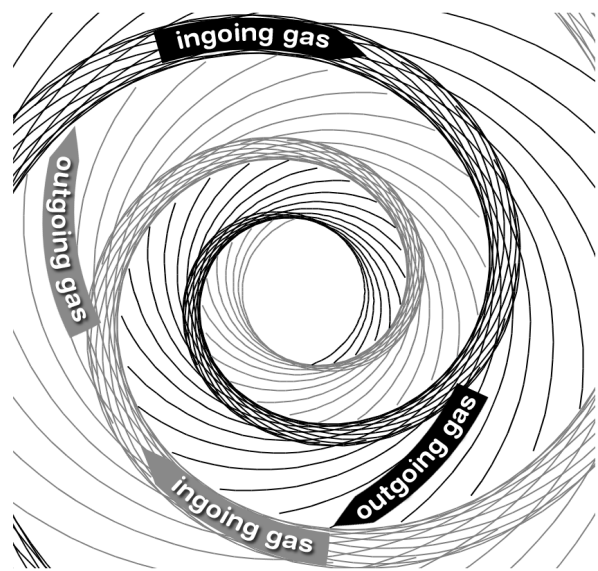

**Figure 19:** Gas motions in a bisymmetric spiral galaxy.

A mechanism for the alignment of orbits with the arms can be understood by plotting the gravitational potential for a spiral galaxy (vertical axis) against the plane of the disc (figure 18). Stellar orbits in such a potential are precisely analogous to the orbits of particles in a frictionless spiral grooved funnel in a uniform gravitational field, for which potential is directly proportional to height. A particle at the highest point of its path, where it is moving least quickly, will tend to fall into a groove and then follow the groove downwards, picking up speed as it goes. Eventually, the particle gains enough momentum to jump free of its groove. It crosses over the next-highest groove (for a bisymmetric spiral), then falls back to a higher point in its original groove. Thus orbits follow the arms on the inward part of the motion, as seen in figure 10 and figure 10, and the tendency for stars to follow the arms reinforces the gravitational potential of the arm. Such a model is insensitive to the shape of the funnel, and could account for the observed frequency of spirals in galaxies with a wide range of sizes and mass distributions.

If gas motions follow a similar pattern to stellar motions, then, in a two-armed spiral, outgoing gas from one spiral arm will collide with ingoing gas in the other arm (figure 19). Gas in the arm will be in turbulent motion, as gas clouds seek to cross in the arm and gain velocity as they approach pericentre. When outgoing gas from one arm meets ingoing gas in another arm, collisions between gas clouds can be expected to create giant molecular clouds, regions of higher density and pressure, and greater turbulence. Pockets of extreme pressure due to collisions and turbulence generate the molecular cores in which new stars form. Because of local variations in the density of ingoing and outgoing gas, and because GMCs and HII regions are expected on orbits crossing the arms, these regions can be expected on an irregular line following the spiral, as was seen in figure 3.

In the main, peaks in the distribution of bright stars coincide with tangencies to the logarithmic spiral (figure 6). However, there is a notable absence of a peak in the Sagitarius sector, and alignment with the 2 kpc and 3 kpc sectors is poor. This may reflect randomness in a relatively small number of massive cloud collisions. The underlying asymmetry in the distribution of bright stars has also been observed in the distribution of red giants (López-Corredoira et al., 2007), but the distribution of red giants does not show the sharp peaks seen here. The reason is that red giants are a more mature population than bright stars, but not so mature that their orbits have achieved alignment with spiral structure. Although gas motions and star formation regions show some adherence to spiral structure, the motions of young stars do not. López-Corredoira et al. interpreted the asymmetry in the distribution of red giants as an indication of a long bar, but we interpret it as indicative of the uneven distribution of star forming regions arising from collisions between massive gas clouds.

In a multi-arm spiral, outgoing gas meeting an arm would have greater mass than ingoing gas in the arm. This would tend to remove gas from the arm. In a two-armed spiral, the gas in the arm would have greater mass. As a result, a two-armed gaseous spiral can be stable, whereas multi-armed gaseous spirals cannot. Outgoing gas would apply pressure to the trailing edge of a spiral arm, and if one gaseous arm advances compared to the bisymmetric position, the pressure due to gas from the other arm will be reduced. At the same time, pressure on the retarded arm due to outgoing gas from the advanced arm will be increased. Thus, since gravity will tend bind the gaseous and stellar spirals, gas motions would tend to preserve the symmetry of two-armed spirals, and account for the observed frequency of grand-design, bisymmetric spiral galaxies.

Digital models have not been successful in showing stable spiral structures. There may be a number of reasons for this. Typically it is only possible to process models with in the order of $10^6$ stars, whereas galaxies may contain in the order of $10^{11}$ stars. In consequence stars are effectively five orders of magnitude more massive than realism dictates, and the gravitational potential is correspondingly less smooth. As a result digital models may cause much greater than realistic scattering. Secondly, digital models may not converge to stability because they do not start from realistic initial conditions. Third, as described above, the stability of the familiar axisymmetric double spiral depends on both gas and stellar motions. Finally the idealised model of spiral structure described above does not include either a bar or a ring. These features can be expected to erode spiral structure from the inside, so that spirals are not the end of galaxy evolution.

## 6 Conclusion

We have reexamined the distributions of neutral gas, GMC's and HII regions. In a plot of $\log(R)$ against angle subtended at the Galactic centre, logarithmic spirals will appear as equidistant parallel lines. This structure is observed for a two-armed logarithmic spiral with pitch angle $5.4 \pm 0.5°$, out to about 12-15 kpc from the Galactic centre, at which point the bisymmetric form breaks down and there appear to be four irregularly spaced arms with much greater pitch angles, continuing outwards beyond 20 kpc.

We also studied the distribution of sources in 2MASS finding peaks in frequency at points consistent with a two-armed spiral with a pitch angle of $5.56 \pm 0.06°$. The bar is clearly identified within the 2MASS data, and has a length $4.2 \pm 0.1$ kpc and an angle $30 \pm 10°$, corresponding to the bulge seen in COBE/DIRBE. Although there is an asymmetry in the distribution of bright stars in 2MASS, we reject the interpretation that this is evidence of a long bar because the distribution contains too much structure. In particular it contains evidence of star formation in spiral arms. We interpret the asymmetry as due to random variations in star forming activity.

The distribution of bright stars becomes ragged at a Galactic radius of 15-20 kpc, in accordance with the notion that star formation is more irregular in the outer regions of the galaxy, but we found no correspondence between peaks in the distribution of sources in 2MASS away from the galactic centre and either the maps of Levine et al. or of Hou et al.

We used two parameters, orbital eccentricity and true anomaly to characterise stream memberships within the local velocity distribution. The majority of orbits of disc stars are broadly aligned with spiral arms, while the dense regions of the streams show strong alignment with the two armed spiral found in the gas distributions. Nearly half of the local population, including the Sun, have orbits broadly aligned with the Orion arm. Our position towards the inside of the arm means that most of these are similarly close to pericentre (Sirius and Alpha Lacertae streams), while a significant number are in the process of rejoining the arm close to apocentre (Pleiades stream). 13.9% of the local population belong to the Hyades stream and have orbits aligned with the Centaurus arm. These are stars crossing the Orion arm on the outward part of their orbit. Three minor streams of stars with higher eccentricity also have orbits aligned with spiral structure. The only significant populations of disc stars not aligned with the arms consist of high velocity stars, and young stars with orbits close to circular motion. The velocity distribution of local stars is thus strongly aligned with a two-armed spiral model, and confirms the finding of Benjamin (2008) that there are only two spiral arms inward of the solar radius.

It has long been difficult to find clear evidence of the spiral structure of the Milky Way. While we interpret gas distributions as indicative of a bisymmetric spiral, the evidence is difficult, if not impossible, to quantify and other commentators have interpreted it differently. The likelihood of finding peaks in the distribution of 2MASS sources at the positions of nine tangencies purely by chance is $3\sigma$ or greater, but this does not exclude the possibility that a part of the distribution of peaks is caused by some unanalysed structure, or by regions of extinction. However, it does not appear to us that the precise alignment of stellar orbits in the solar neighbourhood can be reasonably explained unless this is a genuine indication of the spiral structure of the Milky Way, and we have found substantial agreement between evidence from different sources and of different types that the Milky Way is a grand-design bisymmetric spiral with pitch angle ~5.5°.

**Acknowledgements**

We thank Martin López-Corredoira for comments leading to a number of improvements.